\documentclass[prl,twocolumn,showpacs,reprint]{revtex4}
\usepackage{amsmath}
\usepackage{graphicx} 
\usepackage{dcolumn }
\usepackage{mathptmx}  
\usepackage{array}
\usepackage[normalem]{ulem}
\usepackage{color}
\begin{document}

\title{Optical Gyrotropy and the Nonlocal Hall Effect in Chiral Charge Ordered TiSe$_2$}

\author{Martin Gradhand$^1$}
\email{m.gradhand@bristol.ac.uk}
\author{Jasper  van Wezel$^{2}$}
\email{vanwezel@uva.nl}
\affiliation{$^1$H.H. Wills Physics Laboratory, University of Bristol, Tyndall Avenue, Bristol, BS8 1TL, UK\\
                  $^2$Institute for Theoretical Physics, IoP, University of Amsterdam, 1090 GL Amsterdam, The Netherlands}

\begin{abstract}
It has been suggested that materials which break spatial inversion symmetry, but not time reversal symmetry, will be optically gyrotropic and display a nonlocal Hall effect. The associated optical rotary power and the suggested possibility of inducing a Kerr effect in such materials, in turn are central to recent discussions about the nature of the pseudogap phases of various cuprate high-temperature superconductors. In this letter, we show that optical gyrotropy and the nonlocal Hall effect provide a sensitive probe of broken inversion symmetry in $1T$-TiSe$_2$. This material was recently found to possess a chiral charge ordered phase at low temperatures, in which inversion symmetry is spontaneously broken, while time reversal symmetry remains unbroken throughout its phase diagram. We estimate the magnitude of the resulting gyrotropic constant and optical rotary power and suggest that $1T$-TiSe$_2$ may be employed as a model material in the interpretation of recent Kerr effect measurements in cuprate superconductors.
\end{abstract}


\maketitle

\begin{figure}
\begin{center}
\includegraphics[width=0.9 \columnwidth]{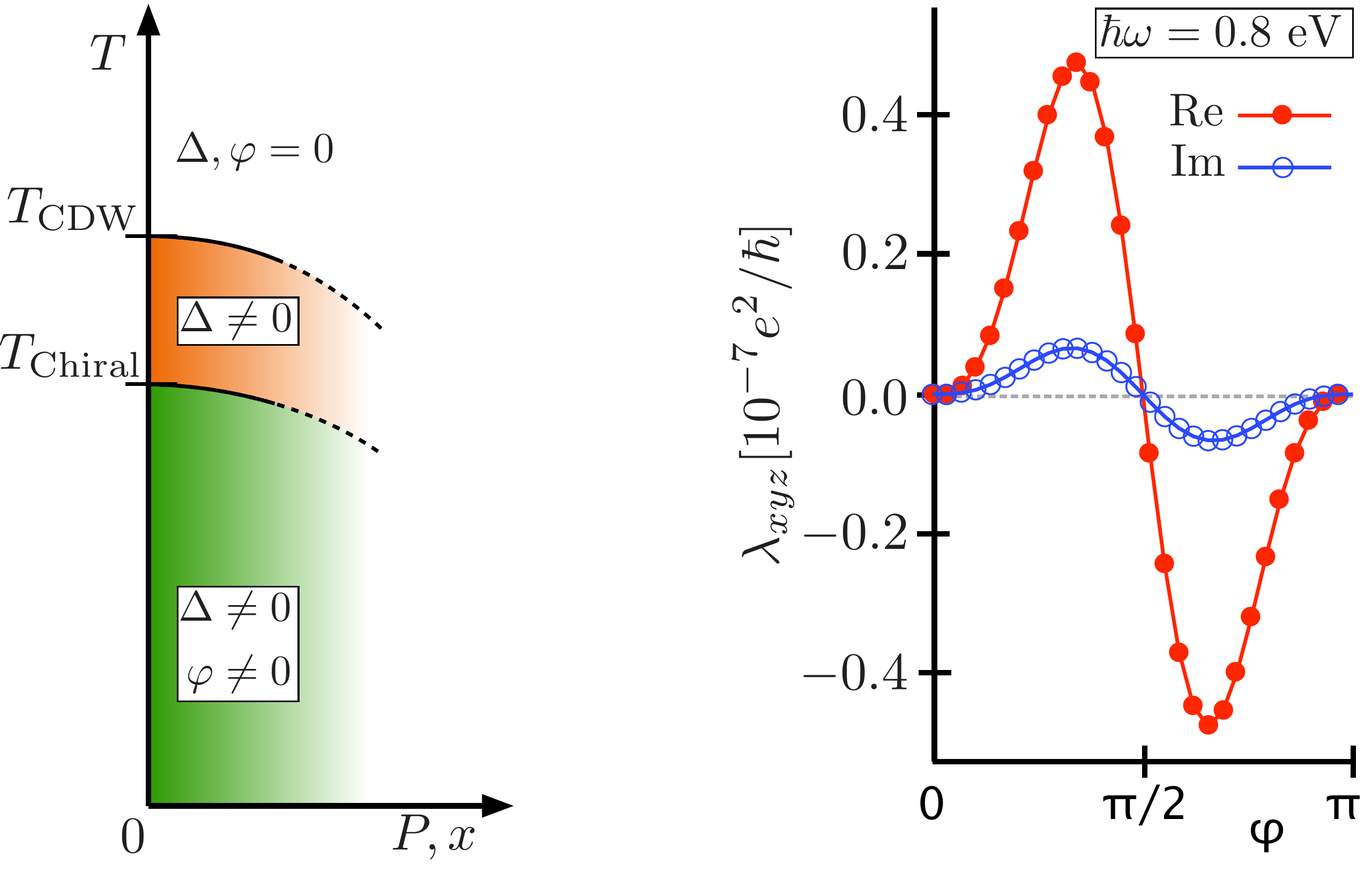}
\end{center}
\caption{(Color online) {\bf Left}: A sketch of the phase diagram of $1T$-TiSe$_2$ as a function of temperature and an external parameter such as pressure or doping. Upon cooling, the material develops a triple-Q, non-chiral charge ordered phase at $T=T_{\text{CDW}}$, before entering into the chiral charge and orbital ordered phase at $T=T_{\text{Chiral}}$.  {\bf Right}: The real and imaginary parts of the nonlocal Hall response as a function of the relative  phase between the three CDW propagation vectors. Nonzero relative phase values indicate the presence of chiral charge order and broken spatial inversion symmetry. \label{Fig.Phase}}
\end{figure}

\emph{Introduction} ---
The measurement of a Kerr effect in the pseudogap phase of several high-temperature superconductors constrains the symmetries that this state may exhibit~\cite{Xia_2008,He_2011,Karapetyan_2012,Hosur_2013}. Although the particular experimental setup used in these studies allows for a non-zero linear response to arise under equilibrium conditions only in the presence of broken time reversal symmetry~\cite{Halperin_1992,Armitage_2014,Fried_2014,Hosur_2015}, it has been argued that the observed optical activity may nonetheless be fundamentally linked to a breakdown of spatial inversion symmetry, related to the presence of charge order~\cite{Hosur_2015}. That it is possible for a charge ordered state to spontaneously break inversion symmetry even in the absence of magnetism or electrostatic polarisation, has only recently become clear, with the discovery of chiral charge order in the low temperature phase of $1T$-TiSe$_2$~\cite{Ishioka_2010,vanWezel_2011,Ishioka_2011,vanWezel_2012,Iavarone_2012,Castellan_2013}. This unexpected emergence of a spiral configuration among the scalar charge density was explained theoretically by the simultaneous presence of orbital order, yielding a vectorial combined order parameter~\cite{vanWezel_2011,Fukutome_1984}. The charge and orbital ordered state in this scenario must arise through a sequence of phase transitions, as indicated schematically in Fig.~\ref{Fig.Phase}, and confirmed experimentally by specific heat, transport and diffraction experiments~\cite{Castellan_2013}.

The chiral charge and orbital order in $1T$-TiSe$_2$ forms an ideal test case for studying the types of phases that have been argued to dominate the optical activity of high-temperature superconductors. It provides an experimentally accessible setting in which charge order breaks spatial inversion symmetry, without the complication of nearby phases with broken time reversal symmetry. In this paper, we show that the chiral order in $1T$-TiSe$_2$ causes the material to be optically gyrotropic. The gyrotropy is evidenced by a non-local Hall effect, and gives rise to non-zero optical rotary power (\emph{i.e.} a Faraday effect at zero magnetic field). Although there cannot be a related non-zero Kerr effect in the presence of time reversal symmetry~\cite{Halperin_1992,Armitage_2014,Fried_2014}, we argue that the presented results indicate that $1T$-TiSe$_2$ can be used as a model system to investigate the relation between the presence of chiral charge order and the observed optical activity of high-temperature superconductors. It allows the effects of non-equilibrium conditions or explicitly broken time reversal symmetry to be studied in a well-understood material within the specific experimental setup used to measure the Kerr effect in cuprate high-temperature superconductors, and can thus be employed to shed light on the interpretation of these measurements.

\begin{figure}
\begin{center}
\includegraphics[width=0.95 \columnwidth]{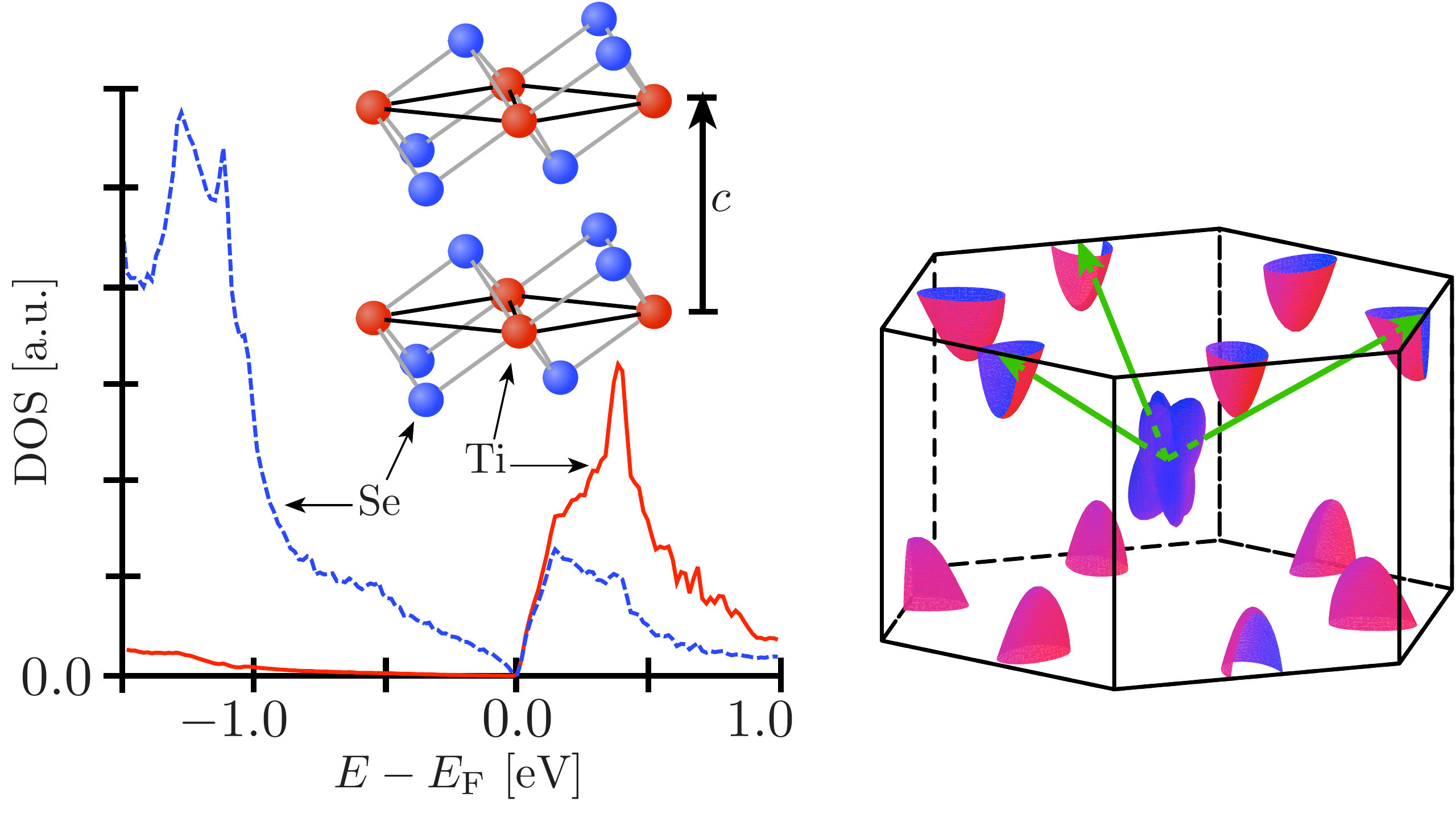}
\end{center}
\caption{(Color online) {\bf Left:} The orbital resolved DOS of the 3-dimensional tight binding model. The inset shows two planes of Ti atoms, each sandwiched between planes of Se atoms, and separated by the lattice constant $c=6$~$\AA$. {\bf Right:} The first Brillouin zone in the normal state, showing the hole pockets around $\Gamma$, and the electron pockets at the $L$ points. Iso-energetic surfaces are shown at energies $E_F+0.08$~eV and $E_F-0.08$~eV for the electron and hole pockets, respectively. The colour code represents the hole (blue) and electron (red) character of the corresponding  states. In addition the three primary CDW Q-vectors are shown.\label{Fig.DOS}}
\end{figure}

\emph{Symmetry Considerations} ---
The normal state of $1T$-TiSe$_2$ consists of quasi two-dimensional layers with hexagonal planes of Ti atoms sandwiched between hexagonal planes of Se atoms. Individual sandwich layers are separated from each other by a Van der Waals gap~\cite{Pehlke_1987,Fang_1997, Goli_2012}. The local coordination of the Ti atoms is octahedral (see inset of Fig.~\ref{Fig.DOS}), resulting in an overall $P\overline{3̄}m1$ space group, characterized by three two-fold rotation axes in the Ti plane, and a three-fold rotation axis perpendicular to that plane. Importantly, the structure preserves both spatial and temporal inversion symmetry. It falls into the point group D$_{3d}$, and accordingly the Laue group $321'$~\cite{Kleiner_1966}. For this Laue group, general group theoretical arguments show that the conductivity tensor must have vanishing off-diagonal elements~\cite{Kleiner_1966,Orenstein_2013}, which rules out the occurrence of a polar Kerr effect. 

This general statement still holds for the non-chiral, triple-Q charge density wave (CDW) forming below T$_{\text{CDW}}$ and is even valid for the chiral state observed below T$_{\text{Chiral}}$ (see left of Fig.~\ref{Fig.Phase}). Although the latter state breaks spatial inversion symmetry, reducing the point group to C$_2$ and the associated Laue group to $21'$, the symmetry class only allows for symmetric contributions to the conductivity tensor, and hence still forbids any polar Kerr effect~\cite{Kleiner_1966,Orenstein_2013}. It was recently pointed out however, that materials without inversion symmetry may be optically gyrotropic~\cite{Orenstein_2013}. They then display a nonlocal Hall effect which is closely related to the optical Kerr effect, and is defined by the linear response:
\begin{align}
j_x=\lambda^G_{xyz}\frac{dE_y}{dz}.
\end{align}      
Here $j_x$ is an electric current in the $x$ direction, flowing in response to the gradient in the $z$ direction of the $y$-component $E_y$ of an applied electric field. The gyrotropic coupling constant $\lambda_{xyz}^G$ can be expressed in terms of the Berry curvature of filled bands as:
\begin{align}\label{Eq_lambda}
\lambda_{xyz}^G &= \frac{e^2}{\hbar}\frac{1}{(1-i \omega\tau_z)^2}\int\limits_{-k_{Fz}}^{+k_{Fz}}dk_z\ \phi(k_z,\omega)v_z(k_z)\tau_z \notag \\
\phi(k_z,\omega) &= \int\limits_0^{k_F(k_z)}d^2k\ \Omega( {\bf k} ,k_z,\omega).
\end{align}
Here, $\Omega( {\bf k},k_z,\omega)$ is the frequency dependent Berry curvature at ${\bf k}=(k_x,k_y)$ within a plane of constant $k_z$. For systems preserving both temporal and spatial inversion symmetry, $\lambda_{xyz}^G$ vanishes, since the Berry curvature vanishes at every point. For a system preserving time reversal symmetry, but not spatial inversions, $\Omega({\bf k},k_z,\omega)$ will be odd under inversion of both ${\bf k}$ and $k_z$. The local Hall conductivity, which is proportional to the average Berry curvature over all filled states, is therefore strictly zero. The nonlocal Hall conductivity arising from the gyrotropic $\lambda_{xyz}^G$ on the other hand, involves the product of two odd functions, $\phi(k_z)$ and $v_z(k_z)$, and is in general nonzero. In the following, we will evaluate the integrals in Eq.~\eqref{Eq_lambda} for a phenomenological tight-binding model of the chiral CDW wave state in $1T$-TiSe$_2$.

\emph{Tight binding model} ---
We solve Eq.~\eqref{Eq_lambda} in the clean limit, approximating $\tau_z \approx 50$~fs to be constant in momentum and the band velocity $v_f\approx 1 \cdot 10^{-6}$~m/s to equal its value at the Fermi energy. In addition, we focus only on the odd part of $\phi(k_z)$, which can contribute to the nonlocal Hall effect. We thus evaluate the expression:
\begin{align}
\lambda_{xyz}^G = \frac{e^2}{\hbar}\frac{2\tau_z v_F}{(1-i \omega\tau_z)^2}\int\limits_{0}^{+k_{Fz}}dk_z\ \left[\phi(k_z,\omega)-\phi(-k_z,\omega)\right].
\end{align}

We employ a three-dimensional, nine band, tight-binding model to describe the electronic structure of the normal state above T$_{\text{CDW}}$. This includes three Ti-$d$ orbitals and three Se-$p$ orbitals each for the upper and lower Se atoms (see the unit cell in Fig.~\ref{Fig.DOS}). This model was previously introduced and described in detail for the case of a single quasi two-dimensional sandwich layer, neglecting the weak couplings between TiSe$_2$ planes~\cite{vanWezel_2010}. In the present case, the dispersion in the $k_z$ direction is crucial for the emergence of the nonlocal Hall effect. We therefore extend the previous model by introducing weak interplane overlap integrals between the upper and lower Se atoms in consecutive layers. For concreteness, we  take $c=6$~$\AA$ as the inter-layer distance. The tight-binding parameters are given in Table~\ref{Tab:para} and result in the orbital-resolved density of states (DOS) presented in the left of Fig.~\ref{Fig.DOS}. The parameters are chosen in order to create a small semiconducting gap ($\sim 0.04$~eV) to avoid spurious effects arising around the Fermi energy at low frequencies. 
\begin{table}[t]
\begin{ruledtabular}
\begin{tabular}{l r l | l r }
$\varepsilon_{\rm Ti}$ & $-0.5625$ &~& $t_{\rm Ti}$ & $0.20$\\ 
$\varepsilon_{\rm Se}$ & $-0.10\phantom{25}$ &~& $t_{\rm Se}$ &  $0.30$ \\
$E_{\text{F}}$ & $0.97\phantom{25}$&~ & $t_{\rm Ti-Se}$ & $1.05$ \\
$\Delta$ & $2.3\phantom{625}$ &~& $t_{\rm Se-Se}$ & $-0.35$ \\ 
$U$ & $0.02\phantom{25}$ &~& $t^{\perp}_{\rm Se-Se}$ & $0.25$
\end{tabular}
\end{ruledtabular}
\caption{The numerical parameters for the tight-binding model of TiSe$_2$ in units of $\text{eV}$. Here, $\varepsilon_{\rm Ti}$ and $\varepsilon_{\rm Se}$ are the on-site energies for Ti and Se atoms. The nearest-neighbour hoppings within the layer are given by $t_{\rm Ti}$, $t_{\rm Se}$, $t_{\rm Ti-Se}$, and $t_{\rm Se-Se}$, while the inter-layer coupling is given by $t^{\perp}_{\rm Se-Se}$. The Fermi energy lies at $E_{\text{F}}$, and the difference in chemical potential between Ti and Se is $\Delta$. The CDW transition is driven by an imposed (mean field) coupling $U$. } 
\label{Tab:para}
\end{table} 

The CDW consists of three components, with propagation vectors connecting the Se states at $\Gamma$ to Ti states at the three inequivalent $L$ points in the first Brillouin zone (see right of Fig.~\ref{Fig.DOS}). The relative phase differences between the three CDW components and their corresponding frozen phonons in real space, are not fixed by the electronic structure alone. Based on free energy arguments, it was shown that two successive phase transitions at $T_{\text{CDW}}$ and $T_{\text{Chiral}}$ separate the high-temperature homogeneous state from a non-chiral CDW phase in which all phase differences vanish, and a chiral CDW at low temperatures~\cite{vanWezel_2011}. In the chiral state, non-zero phase differences imply a breakdown of spatial inversion symmetry, which in turn allows for the presence of optical gyrotropy.

The three CDW components depicted in the right of Fig.~\ref{Fig.DOS} do not form a closed set by themselves. We therefore also consider all four higher harmonics, given by Q$_4=$Q$_1$+Q$_2$, Q$_5=$Q$_1$+Q$_3$, Q$_6=$Q$_2$+Q$_3$, and Q$_7=$Q$_1$+Q$_2$+Q$_3$. All together, this makes the tight-binding Hamiltonian a 72$\times$72 matrix, based on eight sectors for the normal state and all Q vectors, with nine orbital entries each. The coupling between different Q sectors is induced by an imposed mean CDW field along the three vectors Q$_1$, Q$_2$, and Q$_3$. To describe the chiral phase, we allow for a nonzero phase difference $\varphi$ between the Q$_1$ and Q$_2$ CDW components, and $-\varphi$ between the Q$_1$ and Q$_3$ components. For the transport calculations presented below, this eigenvalue problem has to be solved on a very dense mesh of roughly $8 \cdot 10^6$ {\bf k}-points in $1/8$ of the normal state Brillouin Zone.

\emph{Results} ---
In the right of Fig.~\ref{Fig.Phase} we present the gyrotropic coupling constant resulting from the tight-binding model, as a function of the relative phase $\varphi$ between the three Q vectors. As expected, the non-chiral phase at $\varphi=0$, which does not break spatial inversion symmetry, does not show any gyrotropic response. For nonzero relative phase difference a nonzero response does develop, which may be roughly described by a skewed sinusoidal form with a maximum around $\varphi \approx 0.34 \pi$. In the following we will fix the phase to this value.
\begin{figure*}
\begin{center}
\includegraphics[width=0.95 \textwidth]{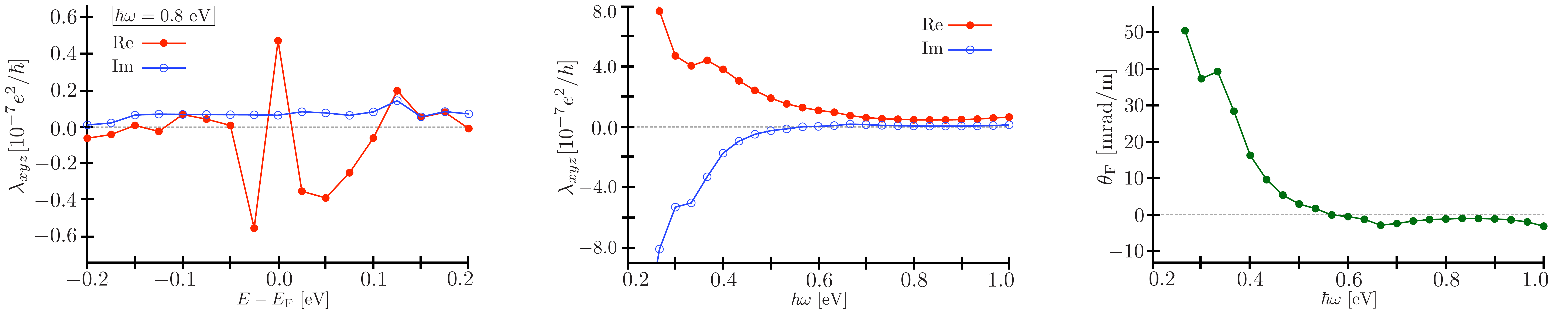}
\end{center}
\caption{(Color online) The real and imaginary parts of the nonlocal Hall conductivity as a function of the chemical potential ({\bf left}) and optical frequency ({\bf middle}). Large fluctuations occur when the chemical potential crosses the semiconducting gap. The frequency dependence is smooth, with $\lambda_{xyz}^G$ increasing towards lower frequencies. {\bf Right:} The optical rotary power as derived from Eq.~\eqref{Eq.:Faraday}. At low frequencies, the rotation of polarization is enhanced, and may be observable.\label{Fig.Faraday}}
\end{figure*}

In the left of Fig.~\ref{Fig.Faraday} we present the real and imaginary parts of the gyrotropic response as a function of chemical potential, for fixed optical frequency $\omega=0.8$~eV. While the imaginary part varies weakly over the considered energy range, the real part shows strong features around the Fermi energy. This is due to the small normal state semiconducting gap, which allows small shifts in chemical potential to result in remarkable changes of the response function. Evidently, this makes any quantitative prediction difficult. 

Fixing the chemical potential and instead varying the frequency of the probing light results in the nonlocal Hall conductivity shown in the middle of Fig.~\ref{Fig.Faraday}. Both the real and imaginary parts are smooth functions of the optical frequency up to $1$~eV. The response drastically increases to lower frequency and the imaginary part changes sign around $\omega=0.56$~eV.

Although the nonlocal Hall conductivity is a clean and direct indication of the presence of broken spatial inversion symmetry and hence chiral charge and orbital order in this system, it may be a difficult quantity to access experimentally. We therefore also present the optical rotary power upon transmission with normal incidence. The transmission geometry limits this application to thin films. The weak Van der Waals bonding between sandwich layers in $1T$-TiSe$_2$ however, allows for relatively straightforward production of thin films with thicknesses ranging from micrometers to nanometers~\cite{Goli_2012,Peng_2014}. Even in single atomic layers, the chiral charge ordered phase is expected to survive~\cite{vanWezel_2012}, and the optical rotary power can therefore provide a direct means of establishing the broken inversion symmetry in this phase. It is calculated using the expression~\cite{Bungay_1993}:
\begin{align}
\theta_F / d  &= -\frac{\mu_0}{2} ~ \textrm{Im}\left[ \gamma^G_{xyz} \omega \right].
\label{Eq.:Faraday}
\end{align}
Here, we changed the notation with respect to Ref.~\onlinecite{Orenstein_2013}, in order to keep SI units and to avoid the need to introduce a nonlocal dielectric tensor. In the right of Fig.~\ref{Fig.Faraday} we present the optical rotary power in $1T$-TiSe$_2$ resulting from Eq.~\eqref{Eq.:Faraday}. It is of the order of $10^{-2}$~rad/m at low frequencies and tracks the imaginary part of the nonlocal conductivity. It diminishes sharply with increasing frequency and changes sign around $0.6$eV, while remaining  of the order of $10^{-3}$~rad/m for higher energies. For films of several micrometers thickness, the rotation of polarization at low frequencies lies at the limit of what is detectable with current state of the art techniques~\cite{Li_2014}.

The closely related optical Kerr effect, or rotation of polarization of reflected light with normal incidence, has previously been suggested as an alternative probe of optical gyrotropy~\cite{Orenstein_2013}, which would not require thin films. It was recently pointed out however, that the standard expression used to calculate the Kerr response~\cite{Bungay_1993}, is incomplete,  as it ignores the variation of gyrotropy at the sample interface~\cite{Halperin_1992,Armitage_2014,Fried_2014,Hosur_2015}. Taking into account these boundary effects, the linear Kerr effect is always identically zero in time reversal symmetric materials under equilibrium conditions. Nevertheless, it has been argued recently that in the Kerr effect measurements of cuprate high-temperature superconductors, the condition of thermal equilibrium may have been weakly violated, allowing for a Kerr response even in the absence of time reversal symmetry breaking~\cite{Hosur_2015}.

In the case of $1T$-TiSe$_2$, a second way to observe a Kerr effect would be to independently satisfy the two symmetry conditions for obtaining a non-zero response~\cite{Hosur_2015}, broken time reversal symmetry and the absence of mirror symmetries. While an applied magnetic field suffices to break time reversal symmetry, it will not affect the mirror symmetries present in the high-temperature atomic lattice. Upon cooling, the lattice symmetries will be spontaneously broken as the chiral transition temperature is traversed, and the onset of a Kerr effect within a magnetic field thus provides a probe for the onset of the chiral charge and orbital ordered state. 

In both the scenario of violated equilibrium conditions, and that of explicitly broken time reversal symmetry, the precise size and shape of the response will depend sensitively on the details of the experimental configuration, and are beyond the scope of the present investigation. The strength and frequency dependence of the cancelled contribution to the standard expression for the Kerr effect on the other hand~\cite{Bungay_1993}, are accessible within the current tight binding model, and are presented in the Supplemental Material. The obtained values in $1T$-TiSe$_2$ are comparable to those in cuprate superconductors, making it plausible that the non-equilibrium and in-field responses of the two classes are likewise of a similar order of magnitude. 

\emph{Conclusions} ---
In summary, we have shown that the breakdown of spatial inversion symmetry and emergence of chirality in the low-temperature charge and orbital ordered phase of $1T$-TiSe$_2$ renders that state optically gyrotropic. The susceptibility to a nonlocal Hall effect serves as a probe for the optical gyrotropy, being zero in both the normal and non-chiral charge ordered phases. In the chiral phase, the nonlocal susceptibility increases with growing relative phase difference between charge density wave components, which can be used as an order parameter for the emerging chirality~\cite{vanWezel_2011,vanWezel_2012}. The gyrotropic response was also found to depend sensitively on the value of the chemical potential, which can be tuned experimentally using various types of intercalants~\cite{Taguchi_1981,Morosan_2006,Morosan_2010,Iavarone_2012}. For non-zero probing frequency, the maximum response is found if the chemical potential lies within a small semiconducting gap, which is close to the condition in the pristine material~\cite{Kidd_2002,Li_2007,May_2011}.

Direct detection of the nonlocal Hall effect is challenging experimentally. The imaginary parts of the nonlocal susceptibility however, is closely tracked by the rotated polarization of normally incident light upon transmission. This optical rotary power is shown to be strongest at low frequencies, and to be of the order of $10^{-2}$~rad/m. For thin films, the optical rotary power may be experimentally accessible, and can serve as a probe of the broken inversion symmetry in $1T$-TiSe$_2$ associated with the onset of chiral order.

The closely related rotation of linear polarization upon reflection of normally incident light, the Kerr effect, was recently shown to be identically zero under equilibrium conditions unless time reversal symmetry is broken~\cite{Halperin_1992,Armitage_2014,Fried_2014,Hosur_2015}. In spite of this result, it has been argued that the observation of a Kerr effect in cuprate high-temperature superconductors may be due to the presence of chiral charge order, without broken time reversal symmetry~\cite{Hosur_2015}. This could be possible if the measurement induced slight non-equilibrium conditions, or if time reversal symmetry was broken explicitly by contaminating magnetic fields or defects. To experimentally test these hypotheses, and thus determine with certainty the symmetries of the pseudogap phase in high-temperature superconductors, a well-understood reference system is required. The chiral charge ordered phase of $1T$-TiSe$_2$ provides such a reference, as it spontaneously breaks inversion symmetry and becomes chiral without breaking time reversal symmetry. At the same time, it can be easily and controllably intercalated with magnetic atoms, subjected to external fields, or driven out of equilibrium. An investigation of the Kerr response in $1T$-TiSe$_2$ thus provides an ideal test bed that can be used to interpret and understand the observed Kerr effect in high-temperature cuprates. 

\begin{acknowledgments}
{\bf Acknowledgements} ---
 MG acknowledges financial support from the Leverhulme Trust via an Early Career Research Fellowship (ECF-2013-538). JvW acknowledges support from a VIDI grant financed by the Netherlands Organisation for Scientific Research (NWO).
\end{acknowledgments}

\end{document}